\documentclass[page-classic]{epl2} 
\usepackage{bm}
\usepackage{amsmath}

\title{Orbital magnetic phase and pure persistent spin current in spin-orbit
coupling mesoscopic rings}
\shorttitle{Orbital magnetic phase and pure persistent spin current}

\author{Guang-Yao Huang \and Shi-Dong Liang\inst{\dag}}
\shortauthor{Guang-Yao Huang \etal}

\institute{
 State Key Laboratory of Optoelectronic Material and Technology and \\
School of Physics and Engineering, Sun Yat-Sen University, Guangzhou,
510275, People's Republic of China\\
\inst{\dag} Corresponding author, Email: stslsd@mail.sysu.edu.cn}


\abstract{
By solving the Rashba model of mesoscopic rings, we give analytically the
ground-state properties of the ring, including the spin polarization, the
persistent charge and spin currents (PCC and PSC). These ground-state
properties can be given based on four kinds of electron numbers in rings.
The effect of the self-inductance of the ring leads to the self-sustained
magnetic flux (SSMF) and the self-sustained PCC and PSC, which break
spontaneously time reversal symmetry to form orbital magnetic phase (OMP).
To tune the spin-orbit coupling strength or electron number of the ring can
induce the phase transition between the OMP and non-OMP. For exact
one-dimensional rings we find the coexistence of the pure PSC and SSMF. This
property of the pure PSC may provide a new scheme to measure the pure PSC.
}

\begin{document}

\maketitle

Spintronics as a new scheme of quantum electronic devices has attracted much
attention, whose central idea is to use the interplay between spin-orbit
(SO) coupling and quantum confinement in semiconductor heterostructures.\cite%
{Prinz,Fabian,Murakami} The structural inversion asymmetry, namely Rashba
effect, leads to intrinsic spin splitting in semiconductor heterostructures.%
\cite{Rashba,Sinova} Many proposals have been put forward for devices based
on spin-dependent transport effect due to the Rashba SO coupling in
low-dimensional systems.\cite{Datta} On the other hand, some quantum phase
and quantum interference effects in low-dimensional systems have attracted
great interests,\cite{Meier} such as the Aharonov-Bohm (AB) oscillation of
the ring conductance induced by the SO coupling,\cite{Morpurgo} and the
persistent charge and spin currents (PCC and PSC) in semiconductor Rashba
rings.\cite{Sheng,Janine,Sun} Although the PCC has been observed
experimentally both in an ensemble of metallic rings\cite{Levy} and single
isolated rings\cite{Mailly}, the experimental investigation of PCC and PSC
in SO coupling semiconductor rings is still a challenging issue for
experimental physicists. Interestingly, the pure PSC was predicted in SO
coupling semiconductor rings,\cite{Sun} and this pure PSC can generate an
electric field, which provides a scheme to measure the pure PSC.\cite{Sun}
Moreover, in principle, the PCC in rings can induce the magnetic flux, which
will modify self-consistently the PCC leading to a self-sustained magnetic
Aharonov-Bohm (AB) flux in rings.\cite{Wohlleben,Zhu} The occurrence of the
self-sustained magnetic AB flux breaks spontaneously time
reversal symmetry to form the orbital magnetic phase (OMP), in which the
self-sustained magnetic AB flux occurs in the ground state even turning off
the external magnetic flux.\cite{Wohlleben,Zhu} However, for semiconductor
Rashba rings both PCC and PSC can occur.\cite{Sheng,Janine,Sun} Whether
still exist OMP in semiconductor Rashba rings and what is relationship
between OMP and PSC? The answers of these questions will provide some
helpful guideline for spintronic devices.

In this letter, we will try to answer above questions. Solving analytically
the Rashba model of mesoscopic rings, we find that the electronic
configuration state can be given analytically based on four kinds of
electron numbers in rings. Thus,all variables, such as PCC, PSC, and the
ground-state energy, can be obtained analytically so that we can give a
clear understanding of whole physical story in semiconductor Rashba rings.
Taking the effect of the self-inductance of the ring into account we will
study the existence of self-sustained magnetic flux (SSMF) and pure PSC and
analyze their relationship. We will also discuss some interesting properties
of the PSC for exact one-dimensional (1D) rings.

Electrons in a two-dimensional(2D) system with a structural inversion
asymmetry can be described by the so-called Rashba model,\cite{Rashba,Sinova}
which involves the SO interaction. For 1D mesoscopic Rashba rings under a
uniform perpendicular magnetic field, the dimensionless Hamiltonian can be
written as \cite{Meijer}
\begin{equation}
\mathcal{H}=\frac{H}{E_{0}}=\left( -i\frac{\partial }{\partial \varphi }%
+f+\sigma _{r}\frac{\bar{\alpha}}{2}\right) ^{2}-\frac{\bar{\alpha}^{2}}{4}
\label{H}
\end{equation}%
where $E_{0}=\frac{\hbar ^{2}}{2m^{\ast }R^{2}}$, where $m^{\ast }$ is the
effective mass of the electron and $R$ is the radius of ring; $f=\frac{\Phi
}{\Phi _{0}}$ is the dimensionless magnetic flux, where $\Phi_{0}$ is the
flux quanta; $\bar{\alpha}=\frac{\alpha }{RE_{0}}$ is the dimensionless
Rashba spin-orbit coupling constant, and $\sigma_{r}=\sigma _{x}\cos \varphi
+\sigma _{y}\sin \varphi$, where $(\sigma_{x},\sigma_{y})$ are the Pauli
matrix and $\varphi$ is the angular variable in the ring. For convenience,
we use the dimensionless form of all variables in the following presentation.

Solving the static Schr\"odinger equation with the Hamiltonian in Eq.(\ref{H}%
) we can give the dimensionless energy spectrum,
\begin{equation}
\mathcal{E}_{n,\sigma }=(n+f+\frac{1}{2}-\frac{\sigma }{2\cos \theta })^{2}-%
\frac{\tan ^{2}\theta }{4}  \label{E}
\end{equation}%
where $\sigma =\pm 1$ for spin up and down; $n=0,\pm1,\pm2\ldots$ label the
electronic states, and $\tan {\theta }=\overline{\alpha }$. The
corresponding eigenvectors can be obtained
\begin{equation}
\psi _{n,\sigma }=\frac{1}{\sqrt{2\pi }}e^{i(n+1/2)\varphi }\left(
\begin{array}{c}
e^{-i\varphi /2}\sin \frac{2\theta +\pi (\sigma +1)}{4} \\
e^{i\varphi /2}\cos \frac{2\theta +\pi (\sigma +1)}{4}%
\end{array}%
\right)  \label{wf}
\end{equation}

Since the energy spectrum in Eq.(\ref{E}) is a periodic function of the
magnetic flux, we may consider the range of the magnetic flux within $|f|<1/2
$. The degenerate points of the energy spectrum arise at the flux $f=-\frac{1%
}{2}(n+n^{\prime }+1-(\sigma +\sigma ^{\prime })\sqrt{1+\bar{\alpha}^{2}}/2)$
for electron states $(n,\sigma )$ and $(n^{\prime },\sigma ^{\prime })$,
which divide the magnetic flux period into six ranges, $(\pm \frac{1}{2},\pm
f_{2}),(\pm f_{2},\pm f_{1}),(\pm f_{1},0)$, where $f_{1}=\frac{1}{2}(\sqrt{%
1+\bar{\alpha}^{2}}-1)$, and $f_{2}=\frac{1}{2}(2-\sqrt{1+\bar{\alpha}^{2}})$%
. For semiconductor rings with its diameter about $50nm$, $\bar{\alpha}%
\approx 1$ and $f_{2}>f_{1}$. When $\bar{\alpha}\geq 1.118$, $f_{2}\leq f_{1}
$.

At zero temperature, the configuration state of electrons is determined by
electron occupation in the energy levels of the ring. For a ring with
electron number $N$ and magnetic flux $f$, we find that the configuration
states can be classified to four kinds of the electron number in above six
ranges of the magnetic flux: case (1) $N=4k+1$ (even number of pairs plus
one extra electron), case (2) $N=4k+2$ (odd number of pairs), case (3) $%
N=4k+3$ (odd number of pairs plus one electron) and case (4) $N=4k+4$ (even
number of pairs), where $k$ is an integer. This property of electronic
configuration states is similar to the non-SO coupling ring.\cite{Loss}
Moreover, we find that there exist a relation of the electron states with
the magnetic flux $f$. For different magnetic fluxes $f$, the electron states $(n(f),\sigma(f))$
satisfy
\begin{equation}
\left\{
\begin{array}{l}
n(-f)+n(f)=-1 \\
\sigma (-f)+\sigma (f)=0%
\end{array}%
\right.  \label{symme}
\end{equation}%
This relation of electron states plays an essential role in giving analytically
the symmetries of the physical quantities such as the charge current, the
spin current and the ground-state energy. We can derive some rules of the
occupation of electron states: (1) $\sum_{n,\sigma }(n(-f)+n(f))=-N$;
(2) $\sum_{n,\sigma }(\sigma (-f)+\sigma (f))=0$; (3) let $\lambda \equiv
\sum_{n,\sigma }(\sigma (-f)n(-f)+\sigma (f)n(f))$, $\lambda =\frac{3}{2}%
N\pm \frac{1}{2}$ for $N=4k+1$ and $N=4k+3$, respectively. $\lambda =2N$ for
$N=4k+2$ and $f_{2}<f<1/2$, and for $N=4k+4$ and $0<f<f_{1}$; $\lambda =N$
for $N=4k+2$ and $0<f<f_{2}$, and for $N=4k+4$ and $f_{1}<f<1/2$. We list
the electron configuration states in the magnetic flux $f>0$ range in the
table I.

\begin{largetable}
\caption{The electronic state configurations in various magnetic flux ranges}
\label{table1}%
\begin{center}
\begin{tabular}{llll}
$N$ & $(\sum_{n\sigma}\sigma,\sum_{n\sigma} n,\sum_{n\sigma} \sigma n)$ &
$(\sum_{n\sigma}\sigma,\sum_{n\sigma} n,\sum_{n\sigma} \sigma n)$ &
$(\sum_{n\sigma}\sigma,\sum_{n\sigma} n,\sum_{n\sigma} \sigma n)$ \\\hline
$4k+1$ & $(1,-\frac{3}{4}(N-1),\frac{3}{4}(N-1))$ &
$(1,-\frac{3}{4}(N-1),\frac{3}{4}(N-1))$ & $(1,-\frac{3}{4}(N-1),\frac{3}{4}(N-1))$\\
$4k+2$ & $(0,-\frac{N}{2},\frac{N}{2})$ & $(0,-\frac{N}{2},\frac{N}{2})$ & $(0,-N,N)$\\
$4k+3$ & $(-1,-\frac{3}{4}(N+1),\frac{3}{4}(N+1))$ &
$(-1,-\frac{3}{4}(N+1),\frac{3}{4}(N+1))$ & $(-1,-\frac{3}{4}(N+1),\frac{3}{4}(N+1))$\\
$4k+4$ & $(0,-\frac{N}{2},N)$ & $(0,-N,\frac{N}{2})$ & $(0,-N,\frac{N}{2})$\\\hline
$f$ & $(0,f_{1})$ & $(f_{1},f_{2})$ & $(f_{2},\frac{1}{2})$ \\
\end{tabular}
\end{center}
{Notes: $f_{1}=\frac{1}{2}(\sqrt{1+\bar{\alpha}^{2}}-1)$, and $f_{2}=\frac{1%
}{2}(2-\sqrt{1+\bar{\alpha}^{2}})$. $k$ is an integer.}
\end{largetable}

It can be seen from the table I that the z-direction projection of the total
spin of the ring in the ground state is $\langle S_{z}\rangle=\frac{\hbar}{2}%
\langle \sum_{n,\sigma}\sigma_{z}\rangle=\pm\frac{\hbar}{2}$ ($0$) for the
odd-(even-)electron ring, respectively. The ground state of rings with
odd-electrons is spin polarized. The spin-up and -down states depend on the
different odd electrons (cases 1 and 3) and the direction of magnetic field.
The ground state of rings with even-electrons is non-spin polarized. This
property may be useful for spintronic devices.

The PCC and PSC can be given by $j_{c(s)}=\sum_{n,\sigma }j_{c(s),n,\sigma }$%
, respectively, where the sum runs over all occupied states. The PCC in each
energy level can be obtained by $j_{c,n,\sigma }=-\partial \mathcal{E}%
_{n,\sigma }/\partial f$, and the PSC in each energy level can be calculated
by $j_{s,n,\sigma }=\frac{1}{2\pi }\int_{0}^{2\pi }2\mathrm{Re}(\psi
_{n,\sigma }^{\dagger }\frac{\{v_{\varphi },s_{z}\}}{2}\psi _{n,\sigma
})d\varphi $, where the velocity operator is $v_{\varphi }=2(-i\frac{%
\partial }{\partial \varphi }+f+\sigma _{r}\frac{\bar{\alpha}}{2})$\cite%
{Janine}.

At zero temperature the PCC and PSC in each energy level can be obtained
\begin{eqnarray}
j_{cc,n,\sigma } &=&-2(n+f+\frac{1}{2}-\frac{\sigma }{2\cos \theta }) \\
j_{sc,n,\sigma } &=&(n+f+\frac{1}{2}-\frac{\sigma }{2\cos \theta })\sigma
\cos \theta   \label{jcs}
\end{eqnarray}%
Using the configuration symmetry Eq.(\ref{symme}) and the electron
occupation rule in the table I, we can prove easily the PCC being an odd
function of flux $f$: $j_{c}(-f)=\sum_{n,\sigma }j_{c,n,\sigma
}(-f)=-j_{c}(f)$, while the PSC is an even function of $f$: $%
j_{s}(-f)=\sum_{n,\sigma }j_{s,n,\sigma }(-f)=j_{s}(f)$. Therefore we may
present the PCC and PSC only in half of the period $0<f<\frac{1}{2}$ . Thus,
we can analytically obtain the PCC,
\begin{equation}
j_{cc,odd}(f)=\left\{
\begin{array}{ll}
-2Nf+\frac{1}{2}(N-3+2\sqrt{\bar{\alpha}^{2}+1}) & \mathrm{for\ }N=4k+1%
\mathrm{;} \\
-2Nf+\frac{1}{2}(N+3-2\sqrt{\bar{\alpha}^{2}+1}) & \mathrm{for\ }N=4k+3%
\mathrm{;}%
\end{array}%
\right.   \label{jco}
\end{equation}%
and%
\begin{equation}
j_{cc,even}(f)=\left\{
\begin{array}{ll}
-N2f & \mathrm{for\ }0<f<f_{1,2}; \\
-N(2f-1) & \mathrm{for\ }f_{1,2}<f<\frac{1}{2};%
\end{array}%
\right.   \label{jce}
\end{equation}%
where $f_{1,2}$ for $N=4k+2$ and $N=4k+4$, respectively. In Figure 1, we
plot the PCC versus the magnetic flux as an example. It can be seen that the
PCC is linear with the magnetic flux and asymmetric at $f=0$. For
odd-electron rings, the PCC has a jump at $f=0$. For even-electron rings the
jump points of PCC shift to $f_{2}$ and $f_{1}$ for the $(4k+2)$- and $(4k+4)
$-electron rings for $\bar{\alpha}=1$, respectively. For $\bar{\alpha}>1.118$%
, the behavior of PCC of odd-electron rings has no qualitative change, but
the jumping points $f_{2}$ and $f_{1}$ of PCC exchange for the $(4k+2)-$ and
$(4k+4)-$number rings. A phase difference of the PCC occurs for the odd- and
even- electron rings. Interestingly, there exist two symmetric non-zero
magnetic fluxes corresponding to zero PCC, which is similar to the numerical result of
the Rashba ring,\cite{Janine} and mesoscopic hard-core boson rings with one impurity.\cite{Zhu}
Another feature of PCC is that the PCC depends
on the SO coupling strength for odd-electron rings, but is independent of
the SO coupling strength for even-electron rings. For given magnetic flux $f$%
, when the SO coupling strength $\bar{\alpha}=\frac{1}{2}\sqrt{[(4f-1)N\pm
3]^{2}-4}$ the PCC vanishes for odd-electron rings, but is still constant
for even-electron rings. This property of PCC is similar to previous
numerical results.\cite{Sun} For the weak SO coupling limit the PCCs in Eqs.(%
\ref{jco}) and (\ref{jce}) are consistent with the PCC of the non-SO
coupling case.\cite{Loss}

\begin{figure}
\onefigure{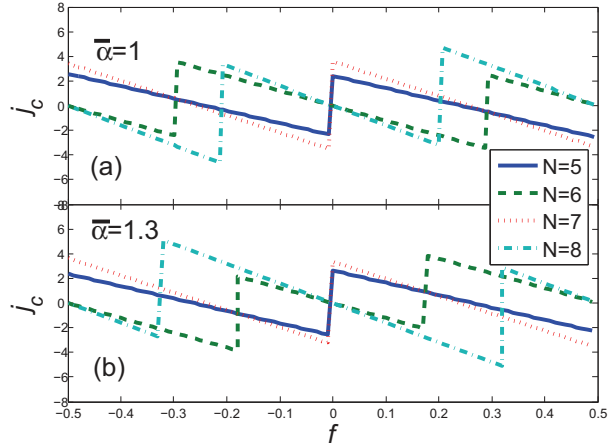}
\caption{The PCC versus the magnetic flux for $\bar{\protect\alpha}=1$ in (a) and for
$\bar{\protect\alpha}=1.3$ in (b).}
\label{fig.1}
\end{figure}

Similarly, the PSC can be also obtained,%
\begin{equation}
j_{sc}(f)=\left\{
\begin{array}{ll}
\frac{2}{\sqrt{\bar{\alpha}^{2}+1}}f+(\frac{3}{2\sqrt{\bar{\alpha}^{2}+1}}%
-1)N-\frac{1}{2\sqrt{\bar{\alpha}^{2}+1}} & \mathrm{for\ }N=4k+1,0<f<\frac{1%
}{2}; \\
-\frac{2}{\sqrt{\bar{\alpha}^{2}+1}}f+(\frac{3}{2\sqrt{\bar{\alpha}^{2}+1}}%
-1)N+\frac{1}{2\sqrt{\bar{\alpha}^{2}+1}} & \mathrm{for\ }N=4k+3,0<f<\frac{1%
}{2}; \\
(\frac{2}{\sqrt{\bar{\alpha}^{2}+1}}-1)N & \left\{
\begin{tabular}{c}
$\mathrm{for\ }N=4k+2,f_{2}<f<\frac{1}{2};$ \\
$\mathrm{for\ }N=4k+4,0<f<f_{1};$%
\end{tabular}%
\right.  \\
(\frac{1}{\sqrt{\bar{\alpha}^{2}+1}}-1)N & \left\{
\begin{tabular}{c}
$\mathrm{for\ }N=4k+2,0<f<f_{2};$ \\
$\mathrm{for\ }N=4k+4,f_{1}<f<\frac{1}{2};$%
\end{tabular}%
\right.
\end{array}%
\right.   \label{js}
\end{equation}

For the odd-electron rings the PSC is linear with the magnetic flux, but for
the even-electron rings the PSC is a constant. We can find that the PSC will
vanish for some SO coupling strengths (see Eq.(\ref{js})), which agrees
qualitatively with the numerical results.\cite{Janine} The PSC versus the
magnetic flux is shown in Fig. 2. We can see that the PSC is linear with the
magnetic flux for odd-electron rings, but have different constants in
different ranges of the magnetic flux for even-electron rings. More
realistically, we may also consider 2D rings. The radial subbands will
induce some additional fine structures of PCC and PSC.\cite{Janine}

In principle, the PCC in a ring can induce a magnetic field and its
corresponding magnetic flux, which interplay with the external magnetic flux
to modify self-consistently the PCC. The dimensionless magnetic energy
induced by the PCC can be written as $\mathcal{E}_{B}=\mu f^{2}$, where $\mu
=\frac{\Phi _{0}^{2}}{2\mathcal{L}E_{0}}$ and the self-inductance coefficient of
the ring $\mathcal{L}$ \cite{Jackson} is $\mathcal{L}=\mu_{0}R[\ln (16R/d)-7/4]$, where $d$
is the diameter of the cross section of the ring. The total energy of the whole system can be
written as
\begin{equation}
\mathcal{E}_{T}=\sum_{n,\sigma }\left[ (n+f+\frac{1}{2}-\frac{\sigma }{2\cos
\theta })^{2}-\frac{\tan ^{2}\theta }{4}\right] +\mu f^{2},  \label{ET}
\end{equation}%
Physically, for a given external magnetic field the stable state of the
whole system should be at the minimum of the total energy $\mathcal{E}_{T}$.
Solving the equation $\frac{\partial \mathcal{E}_{T}}{\partial f}=0$, we can
obtain the magnetic flux of the ring in the ground state for given electron
number $N$.
\begin{equation}
f_{odd}^{s}=\left\{
\begin{array}{ll}
\frac{1}{4(N+\mu )}(N-3+2\sqrt{\bar{\alpha}^{2}+1}) & \mathrm{for\ }N=4k+1%
\mathrm{;} \\
\frac{1}{4(N+\mu )}(N+3-2\sqrt{\bar{\alpha}^{2}+1}) & \mathrm{for\ }N=4k+3%
\mathrm{;}%
\end{array}%
\right.  \label{fo}
\end{equation}%
and%
\begin{equation}
f_{even}^{s}=\frac{N}{2(N+\mu )}\mathrm{\ for\ }N\text{ }\mathrm{even,with}%
\text{ }f_{1,2}<f_{even}^{s}<\frac{1}{2};  \label{fe}
\end{equation}%
where $f_{1,2}$ for $N=4k+2$ and $N=4k+4$, respectively. Moreover, we can
demonstrate $\frac{\partial ^{2}\mathcal{E}_{T}}{\partial f^{2}}=2(N+\mu )$,
which satisfies the condition $\left( \frac{\partial ^{2}\mathcal{E}_{T}}{%
\partial f^{2}}\right) |_{f=f_{s}}>0$ for $f$\ in the whole period. It
implies that the whole system is stable for the magnetic flux $%
f_{odd(even)}^{s}$ in Eqs.(\ref{fo}) and (\ref{fe}). Thus, when $%
f_{odd(even)}^{s}\neq 0$, there exists a SSMF of the ring even though the
external magnetic field is turned off after the system reaches stable
states. It implies that the time reversal symmetry is broken spontaneously
to form the orbital magnetic phases (OMP) like the metallic rings.\cite%
{Wohlleben,Zhu} The SSMF may be regarded as the order parameter of OMP.

\begin{figure}
\onefigure{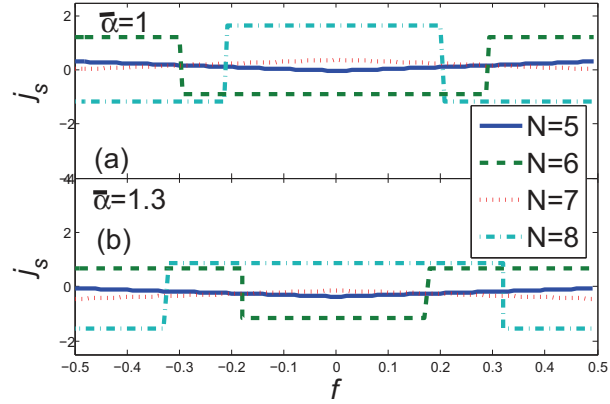}
\caption{The PSC versus the magnetic flux for $\bar{\protect\alpha}=1$ in (a) and for
$\bar{\protect\alpha}=1.3$ in (b).}
\label{fig.2}
\end{figure}

The SSMF we give in Eqs.(\ref{fo}) and (\ref{fe}) depends on three
parameters of the ring, the self-inductance $\mathcal{L}$ (inside $\mu $),
the electron number $N$ and the spin-orbit coupling strength $\bar{\alpha}$.
For odd-electron number rings, when $\bar{\alpha}=\sqrt{\frac{1}{4}%
(N\mp3)^{2}-1}$, $f^{s}_{odd}=0$. This means that tuning the SO coupling
strength can induce the phase transition between the orbital magnetic and
non-orbital magnetic orders. This provides a possible way to tune the OMP phase transition
because the SO coupling strength can be tuned by applied
electric field in semiconductor heterostructures. It should be emphasized
that the SSMF corresponds to a spin polarized state for odd-electron number
rings, but corresponds to non-spin polarized state for even-electron number
rings. This is quite different from the SSMF in metallic mesoscopic rings
without SO coupling,\cite{Wohlleben,Zhu} in which the SSMF always
corresponds to non-spin polarized state.

Substituting the SSMF in Eqs.(\ref{fo}) and (\ref{fe}) into Eqs. (\ref{jco})
and (\ref{jce}), we can obtain the self-sustained PCC, $j_{cc}^{s}=2\mu
f_{odd(even)}^{s}$. Similarly, the self-sustained PSC for odd-electron rings
can be given,
\begin{equation}
j_{sc}^{s}=\frac{1}{2\sqrt{\bar{\alpha}^{2}+1}}(3N-\frac{3\pm \mu }{N+\mu })+%
\frac{1}{N+\mu }-N,  \label{jss}
\end{equation}%
where '$+$'\ for $N=4k+1$ and '$-$'\ for $N=4k+3$ in the range $0<f<\frac{1}{%
2}$. For even-electron rings, the self-sustained PSC is still constant same
to Eq.(\ref{js}).

For the 1D limit $d\rightarrow 0$, the self-inductance $\mathcal{L}%
\rightarrow \infty $, and $\mu \rightarrow 0$, the classical magnetic field
energy can be ignored.\cite{Zhu} The self-sustained PCC vanishes $j_{cc}^{s}=0
$, but the SSMF is still finite, which is similar to the self-sustained
AB flux in mesoscopic hard core boson ring.\cite{Zhu}
Interestingly, the self-sustained PSC becomes,
\begin{equation}
j_{sc}^{s}=(\frac{3}{2\sqrt{\bar{\alpha}^{2}+1}}-1)(N-\frac{1}{N}),
\label{selfPSC}
\end{equation}%
for odd-$N$. This is a self-sustained pure PSC, namely there only exists the
spin current without charge current, which may be also regarded as a
spontaneous time reversal symmetry breaking. This self-sustained pure PSC
gives some novel properties: (1) it accompanies a nonzero magnetic flux even
though it is pure PSC without PCC; (2) it is sensitive to electron number $N$
and the SO coupling strength; (3) it still survives even for weak SO
couplings, but when $\bar{\alpha}=\sqrt{5}/2$, the self-sustained pure PSC
vanishes. These properties of PSC are quite different from the pure PSC
predicted by Sun \textit{et.al.}\cite{Sun}. Theoretically, it is
interesting to demonstrate the existence of the pure PSC.
It has been found that the pure PSC can induce an electric field.\cite{Sun,Hirsh}
One has proposed a scheme to demonstrate the existence of the pure PSC by measuring this electric
field.\cite{Meier,Schutz} The self-sustained pure PSC in Eq.(\ref{selfPSC}) coexists
with the SSMF except $\bar{\alpha}=\sqrt{5}/2$. Thus once we can measure the magnetic field
associated with the SSMF, we can demonstrate the existence of pure PSC as long as $\bar{\alpha}\neq\sqrt{5}/2$.
For the ring of $500nm$ diameter, the magentic field of SSMF is estimated approximately as $6$ $Gauss$.

For the Dresselhaus model, since the energy spectrum has on qualitatively
difference from the Rashba, all properties we obtain from Rashba model
should not have qualitatively change for the Dresselhaus model.\cite{Sheng}

In summary, the theoretical study of the Rashba model of mesoscopic rings
reveals some quantum phenomena in semiconductor rings. The spin polarization
in the ground state depends on the odd-even electron number of the ring, but
the PCC and PSC can be given based on four kinds of electron numbers in the
ring. The effect of the self-inductance of the ring leads to the SSMF and
the self-sustained PCC and PSC, which breaks spontaneously time reversal
symmetry to form OMP. The phase transition between OMP and non-OMP
can be induced by tuning the SO coupling strength or the electron number of
the ring. For exact 1D rings we find the coexistence of the pure PSC and
SSMF, which may provide a theoretical scheme to measure the pure PSC.

\acknowledgments

The author gratefully acknowledges the financial supports of the projects
from the National Natural Science Foundation of China (Grants No.10774194, U0634002),
National Basic Research Program of China (973 Program: 2007CB935501), and
Advanced Academic Research Center of Sun Yat-Sen university (06P4-3).


\end{document}